\documentclass[12pt]{article}
\usepackage{amssymb,amsmath,epsfig}
\allowdisplaybreaks

\begin{document}
\title{\bf Influence of Electric Charge and Modified Gravity on Density Irregularities}

\author{M. Zaeem Ul Haq Bhatti \thanks{mzaeem.math@pu.edu.pk} and Z. Yousaf
\thanks{zeeshan.math@pu.edu.pk}\\
Department of Mathematics, University of the Punjab,\\
Quaid-i-Azam Campus, Lahore-54590, Pakistan.}

\date{}

\maketitle
\begin{abstract}
This work aims to identify some inhomogeneity factors for plane
symmetric topology with anisotropic and dissipative fluid under the
effects of both electromagnetic field as well as Palatini $f(R)$
gravity. We construct the modified field equations, kinematical
quantities and mass function to continue our analysis. We have
explored the dynamical quantities, conservation equations and
modified Ellis equations with the help of a viable $f(R)$ model.
Some particular cases are discussed with and without dissipation to
investigate the corresponding inhomogeneity factors. For
non-radiating scenario, we examine such factors with dust, isotropic
and anisotropic matter in the presence of charge. For dissipative
fluid, we investigate the inhomogeneity factor with charged dust
cloud. We conclude that electromagnetic field increases the
inhomogeneity in matter while the extra curvature terms make the
system more homogeneous with the evolution of time.
\end{abstract}

\section{Introduction}

The inclusion of higher order curvature invariants in the action for
the modifications of general relativity (GR) have a long primordial
history. An alternative approach hypothesizes that GR is accurate
only on small scales and has to be generalize on large/cosmological
distances. The early struggle was mostly due to the scientific
curiosity to understand the newly proposed theory and to give some
alternative to dark energy model. However, new motivations came from
some theoretical aspects of its physics which revived the study of
higher order gravity theories \cite{z2}. To initiate with, there are
various techniques and proposal for modified gravity to deviate from
GR. The $f(R)$ theories of gravity \cite{z3} are the straightforward
generalization of the Einstein Hilbert action, in which the Ricci
scalar $(R)$ becomes a generic function of $R$. In fact, it is
relatively simple and compelling alternative to GR, from which some
important results have already been obtained in the literature.

It is worth mentioning that one can apply two variational principles
to derive $f(R)$ field equations from the modified form of
Einstein-Hilbert action. One is the standard metric variation while
the second one is dubbed as the Palatini variation in which the
connection and metric dealt as independently. More precisely, one
has to vary the action with respect to both metric and connection in
such a manner that the matter action does not depend upon
connection. Accordingly, there would be two versions of $f(R)$
gravity corresponding to which the variational formalism is
explored. Here, the Einstein-Hilbert action can be modified through
its gravitational part in order to discuss the $f(R)$ theory of
gravity as \cite{1}
\begin{equation}\nonumber
S_{f(R)}=\frac{1}{2\kappa}\int d^4x\sqrt{-g}f(R)+S_M,
\end{equation}
where $\kappa,~S_M$ and $f(R)$ are coupling constant, matter action
and a non-linear Ricci function, respectively. Applying the
variation with metric ($g_{\alpha\beta}$) and connection
($\Gamma^\rho_{\alpha\beta}$) in the above action, respectively, one
can formulate the following couple of equations of motion as
\begin{eqnarray}\label{1}
&&f_R(\breve{R}){\breve{R}}_{\alpha\beta}-[g_{\alpha\beta}f(\breve{R})]/2
={\kappa}T_{\alpha\beta},\\\label{2} &&
\breve{\nabla}_\mu(g^{\alpha\beta}\sqrt{-g}f_R(\breve{R}))=0.
\end{eqnarray}
By taking the trace of Eq.(\ref{1}), we can constitute an analogy
between $T\equiv g^{\alpha\beta}T_{\alpha\beta}$ and $R \equiv
R(\Gamma)$ as
\begin{equation}\label{3}
Rf_R({R})-2f({R})={\kappa}T,
\end{equation}
which describes the reliance of Ricci scalar on $T$. To examine a
consistent Palatini $f(R)$ gravity with any other classical theory,
we have to deal with only such situations where the solution of the
above equation exists. With present cosmological value of Ricci
invariant, i.e., $R=\tilde{R}$, Eq.(\ref{3}) leads to the covariant
conservation of metric thereby fixing $\Gamma^\rho_{\alpha\beta}$ to
Levi-Civita. Consequently, for vacuum cases, Eq.(\ref{1}) turns out
to be
\begin{equation}\label{4}
\breve{R}_{\alpha\beta}-\Lambda(\tilde{R})g_{\alpha\beta}=0,
\end{equation}
where $\breve{R}_{\alpha\beta}$ is called metric Ricci tensor of
$g_{\alpha\beta}$ and $\Lambda(\tilde{R})={\tilde{R}}/{4}$. This
theory would lessen to GR in the presence/absence of cosmological
constant depending on a viable $f(R)$ model. One can obtain a single
expression for field equations in Palatini $f(R)$ formalism by
substituting $\Gamma^\sigma_{\alpha\beta}$ from Eq.(\ref{2}) in
terms of $g_{\alpha\beta}$ as follows
\begin{align}\nonumber
&\frac{1}{f_R}\left(\breve{\nabla}_\alpha\breve{\nabla}_\beta-g_{\alpha\beta}
\breve{\Box}\right)f_R+\frac{1}{2}g_{\alpha\beta}\breve{R}+\frac{\kappa}{f_R}
T_{\alpha\beta}+\frac{1}{2}g_{\alpha\beta}\left(\frac{f}{f_R}-R\right)
\\\label{5}
&+\frac{3}{2f_R^2}\left[\frac{1}{2}g_{\alpha\beta}(\breve{\nabla}
f_R)^2-\breve{\nabla}_\mu f_R\breve{\nabla}_\beta
f_R\right]-\breve{R}_{\alpha\beta}=0,
\end{align}
which can be written in an alternative form as
\begin{equation}\label{6}
\breve{G}_{\alpha\beta}=\frac{\kappa}{f_R}(T_{\alpha\beta}
+{\mathcal{T}_{\alpha\beta}}),
\end{equation}
here
\begin{eqnarray*}\nonumber
{\mathcal{T}_{\alpha\beta}}&=&\frac{1}{\kappa}\left(\breve{\nabla}_\alpha\breve{\nabla}_
\beta-g_{\alpha\beta}\breve{\Box}\right)f_R+\frac{f_R}{2\kappa}g_{\alpha
\beta}\left(\frac{f}{f_R}-R\right)\\\nonumber
&+&\frac{3}{2{\kappa}f_R}\left[\frac{1}{2}g_{\alpha\beta}(\breve{\nabla}
f_R)^2-\breve{\nabla}_\alpha f_R\breve{\nabla}_\beta f_R\right]
\end{eqnarray*}
is the effective energy-momentum tensor in Palatini $f(R)$ terms
describing modified gravitational contribution while
$\breve{G}_{\alpha\beta}~\equiv~\breve{R}_{\alpha\beta}-\frac{1}{2}g_{\alpha\beta}\breve{R},\quad
\breve{\Box}=\breve{\nabla}_\alpha\breve{\nabla}_\beta
g^{\alpha\beta}$, where $\breve{\nabla}_\alpha$ shows covariant
derivative with respect to Levi-Civita connection. It is interesting
to note that $f_R$ and $f$ are functions of
$R(\Gamma)~\equiv~g^{\alpha\beta}R_{\alpha\beta}(\Gamma)$. If one
disregard the supposition that matter action is independent of
connection, then a new version of $f(R)$ gravity is observed, called
metric-affine $f(R)$ gravity, which have both Palatini and metric
$f(R)$ on its usual limits. The viability criteria for any
gravitational theory include \cite{z3, z4} stability, correct
Newtonian and post-Newtonian limits, a correct cosmological
dynamics, cosmological perturbations compatible with large scale
structures and cosmic microwave background and the absence of
ghosts. Many interesting results emerge from $f(R)$ gravity as they
predict the early universe inflation and have a well-posed Cauchy
problem. Nojiri and Odintsov \cite{2} studied various modified
gravity models as an alternative to dark energy. They investigate
that inhomogeneous terms are originated due to modified gravity
model of the universe. Guo and Joshi \cite{3} examined the collapse
of spherical star due to Starobinsky $R^2$ model within the
framework of $f(R)$ gravity. Here, we would like to discuss the
inhomogeneities/irregularities which emerge in the energy density
due to Palatini version of $f(R)$ gravity.

Anisotropic effects are leading paradigms in the description of
evolutionary mechanisms of stellar collapsing models. It is an
established fact the properties of anisotropic models may differ
drastically in contrast with the isotropic spheres. Nguyen and
Pedraza \cite{4} investigated anisotropic spherical compact model
and deduced that anisotropic effects make the system dissipative
with the evolution of time. Leon and Sarikadis \cite{5} investigated
impact of anisotropy in the framework of modified gravity and
concluded different cosmological  behaviors in the geometry as
compare to isotropic scenarios. Cosenza et al. \cite{6} figured out
the role of anisotropy on radiating fluid spheres. Maartens et al.
\cite{7} analyzed the anisotropic evolution of the universe during
an intermediate transient regime of inflationary expansion.

The anisotropic picture in relativistic fluid configurations can be
achieved by many interconnected phenomena like the existence of
strong electric and magnetic interactions \cite{8}. A great deal of
attention has also been given to the interaction of electromagnetic
and gravitational fields. However, a general harmony exist among the
relativists that physical objects with a large amount of electric
charge does not exist in nature. This view of thought have been
challenged by many researchers and a variety of work have been
carried out with this background. Ghezzi \cite{10} explored some
analytical models of isotropic spherical star in the presence of
electromagnetic field in which the charge density is proportional to
the rest mass density. He found that the radius of charged stars is
larger as compared to the uncharged ones. Varela et al. \cite{11}
solved the Einstein-Maxwell field equations for self-gravitating
anisotropic spherical system numerically and link their findings
with the models of dark matter including massive charged particles
as well as charged strange quark stars. The impact of
electromagnetic field and other matter variables on the evolutionary
behavior of collapsing relativistic self-gravitating systems in
cosmos have been investigated in \cite{z5} and \cite{znew1}.

A system begins collapsing once it experiences an inhomogeneous
stellar state. Penrose and Hawking \cite{12} explored irregularities
in the energy density of spherical relativistic stars by means of
Weyl invariant. Herrera \textit{et al.} \cite{13} discussed the role
of density inhomogeneities on the structure and evolution of
spherically anisotropic objects. Herrera et al. \cite{ya29} did a
systematic study on structure formation of self-gravitating compact
stars by means of some scalar functions (trace and trace-free parts)
obtained from splitting of the Riemann tensor. These scalars are
associated with electric, magnetic as well as second dual of the
Riemann tensor and have eventual relationship with fundamental
properties of the matter configuration \cite{ya30}. The
inhomogeneity in the universe can be linked with the dipole
anisotropy as found by Planck \cite{14}. Herrera \cite{18}
investigated different physical factors responsible for the
emergence of inhomogeneities in an initial regular spherical
collapsing distributions. Sharif and Yousaf \cite{z1} described
stability of regular energy density in planar matter distribution by
taking into account three parametric model form in Palatini $f(R)$
gravity.

The inhomogeneous models can also be used to discuss the SN-data
\cite{15}. Geng and L\"{u} \cite{16} presented a class of models
describing the isotropic expansion for inhomogeneous universe. The
unresolved issues of the dark energy/dark matter on the homogeneity
of the collapsing compact star is still a matter of
interest for relativists. We will address two main related problems in this paper:\\

\noindent\textbf{1.} We explore inhomogeneity factors for plane
symmetric compact object and discuss it with some particular cases
by
increasing the complexity in the matter distribution. \\
\textbf{2.} The role of Palatini $f(R)$ dark source terms through a
viable $f(R)$ model as well as
electromagnetic field effects will be analyzed.\\

This paper is outlined in following manner. In the next section, we
deduce field equations coupled with the source in $f(R)$ gravity
under the influence of electromagnetic field. Section \textbf{3}
investigates the dynamical as well as evolution equations for the
systematic analysis of inhomogeneity factors. In section \textbf{4},
we formulate the irregularity factors with some particular cases of
dissipative and non-dissipative matter field. Finally, we conclude
our main findings in the last section.

\section{$f(R)$ Gravity Coupled to Matter Source}

We choose a non-static planar geometry for the construction of our
systematic analysis as \cite{17}
\begin{equation}\label{7}
ds^2_-=-A^2(t,z)dt^{2}+B^2(t,z)\left(dx^{2}+dy^2\right)+C^2(t,z)dz^2,
\end{equation}
while it is filled with dissipative fluid by means of diffusion
(heat) as well as free streaming (null radiation) approximations
having anisotropic pressure in the interior. Such matter fields are
described by the energy-momentum tensor as follows
\begin{align}\label{8}
T_{\alpha\beta}=(P_{\bot}+\mu)V_\alpha
V_\beta+q_{\beta}V_\alpha+{\varepsilon}l_\alpha l_\beta+P_\bot
g_{\alpha\beta}+(P_z-P_\bot)\chi_\alpha\chi_\beta+q_{\alpha}V_\beta,
\end{align}
where $\varepsilon,~\mu,~P_{\bot},~P_r$ and $q_{\beta}$ are the
radiation density, energy density, different stress components and
heat flux vector, respectively.

In comoving coordinate system, the unit four vector
$l^\beta=\frac{1}{A}\delta^{\beta}_{0}+\frac{1}{C}\delta^{\beta}_{3}$,
radial four vector, i.e.,
$\chi^{\beta}=\frac{1}{C}\delta^{\beta}_{3}$ as well as four
velocity vector $V^{\beta}=\frac{1}{A}\delta^{\beta}_{0}$, satisfies
the following relations
\begin{align}\nonumber
&V^{\alpha}V_{\alpha}=-1,\quad\chi^{\alpha}\chi_{\alpha}=1,
\quad\chi^{\alpha}V_{\alpha}=0,\\\nonumber &V^\alpha q_\alpha=0,
\quad l^\alpha V_\alpha=-1, \quad l^\alpha l_\alpha=0.
\end{align}
The expansion rate of matter configuration under Palatini $f(R)$
background is defined by the scalar as
\begin{align}\label{9}
\Theta_{P}=V_{\alpha;\beta}V^{\beta}=\frac{2}{A}\left(\frac{\dot{B}}{B}
+\frac{\dot{f_R}}{f_R}+\frac{\dot{C}}{2C}\right),
\end{align}
where dot indicates the operator $\frac{\partial}{\partial{t}}$. The
shear scalar for planar in the framework of GR yields \cite{18}
\begin{align}\label{10}
9\sigma^2=\frac{9}{2}\sigma^{\mu\nu}\sigma_{\mu\nu}={W^2_{GR}},\quad
\textrm{with}\quad
{W_{GR}}=\frac{1}{A}\left(\frac{\dot{C}}{C}-\frac{\dot{B}}{B}\right).
\end{align}
Using Eqs.(\ref{9}) and (\ref{10}), we can determine a relation
between expansion and shear as follows
\begin{align}\label{11}
{W_{GR}}=\Theta_P-\frac{3\dot{C}}{AC}-\frac{2\dot{f_R}}{Af_R}.
\end{align}

The stress-energy tensor describing the electromagnetic field and
satisfying the Maxwell field equations, i.e.,
$F^{\alpha\beta}_{~~;\beta}={\mu}_{0}J^{\alpha},\quad
F_{[\alpha\beta;\gamma]}=0,$ is defined as
\begin{equation*}\label{5}
E_{\alpha\beta}=\frac{1}{4\pi}\left(F^{\gamma}_{\alpha}F_{\beta\gamma}-
\frac{1}{4}F^{\gamma\delta}F_{\gamma\delta}g_{\alpha\beta}\right),
\end{equation*}
where $F_{\alpha\beta}=-\phi_{\alpha,\beta}+\phi_{\beta,\alpha}$ is
the Maxwell strength tensor where $\phi_\beta$ describes four
potential. Here $J_\alpha$ and $\mu_0=4\pi$ represent four current
and magnetic permeability, respectively. The four potential and four
current are $\phi^{\alpha}={\phi} {\delta^{\alpha}_{0}},\quad
J^{\alpha}={\xi}V^{\alpha},$ under comoving coordinate system while
$\phi,~\xi$ are functions of $t$ and $z$ representing the scalar
potential and charge density, respectively. The non-zero components
of the Maxwell field equations yields the following couple of
equations as
\begin{eqnarray}\label{7a}
\frac{\partial^2\phi}{\partial{z}^2}-\left(\frac{A'}{A}+\frac{C'}{C}
-\frac{2B'}{B}-\frac{2F'}{F}\right)\frac{\partial\phi}{\partial{z}}&=&\sigma{\mu}_{0}AC^2,
\\\label{8a}
\frac{\partial^2\phi}{\partial{t}\partial{z}}-\left(\frac{\dot{A}}{A}
+\frac{\dot{C}}{C}-\frac{2\dot{B}}{B}-\frac{2\dot{F}}{F}\right)\frac{\partial\phi}
{\partial{z}}&=&0.
\end{eqnarray}
Here prime indicates $z$ differentiation. Integration of
Eq.(\ref{7a}) with respect to $z$ provides
\begin{eqnarray}\label{9a}
{\phi}'=\frac{sAC}{F^2B^{2}},\quad\textmd{where}\quad s
=\mu_0\int^z_{0}{\sigma}F^2C{B^2}dz,
\end{eqnarray}
which equivalently satisfies Eq.(\ref{8a}). The non-vanishing
components of the electromagnetic stress tensor turns out to be
\begin{equation}
E_{00}=\frac{s^2}{8\pi A^2B^4},\quad E_{11}=\frac{s^2}{8\pi
B^6}=E_{22},\quad E_{33}=-\frac{s^2}{8\pi B^4C^2}.
\end{equation}

The field equations in the framework of Palatini $f(R)$ gravity
corresponding to planar geometry leads to
\begin{align}\nonumber
&\frac{\kappa}{f_{R}}\left[A^{2}({\mu}+{\varepsilon})+\frac{s^2}{8\pi
A^2B^4}-\frac{A^2}
{\kappa}\left\{\frac{f_{R}'}{C^2}\left(\frac{C'}{C}+\frac{{f'_R}}{4f_R}-\frac{2B'}{B}
\right)-\frac{f_R}{2}\left(R-\frac{f}{f_R}\right)\right.\right.\\\nonumber
&\left.\left.-\frac{f_{R}''}
{C^2}+\left(\frac{\dot{C}}{C}+\frac{9\dot{f_R}}{4f_R}+\frac{2\dot{B}}{B}
\right)\frac{\dot{f_{R}}}{A^2}\right\}\right]=\left(\frac{\dot{B}}{B}\right)^2+\frac{
2\dot{C}\dot{B}}{CB}+\left\{\frac{B'}{B}\left(\frac{2C'}{C}-
\frac{B'}{B}\right)\right.\\\label{12}
&\left.-\frac{2C''}{C}\right\}\left(\frac{A}{C}\right)^2,\\\nonumber
&\frac{\kappa}{f_{R}}\left[CA(q+{\varepsilon})-\frac{1}{\kappa}
\left(\dot{f_{R}'}-\frac{5}{2}\frac{\dot{f_R}f'_R}{f_R}-\frac{\dot{C}f_{R}'}{C}-\frac{A'\dot{f_{R}}}{A}
\right)\right]=2\left(\frac{\dot{B'}}{B}-\frac{A'\dot{B}}{BA}\right.\\\label{13}
&\left. -\frac{B'\dot{C}}{BC}\right),\\\nonumber
&\frac{\kappa}{f_{R}}\left[{P_{\bot}}B^{2}+\frac{s^2}{8\pi
B^6}+\frac{B^2}{\kappa}
\left\{\frac{\ddot{f_{R}}}{A^2}-\frac{f_R''}{C^2}+\left(\frac{
\dot{B}}{B}-\frac{\dot{f_R}}{4f_R}-\frac{\dot{A}}{A}+\frac{\dot{C}}{C}\right)\frac{
\dot{f_{R}}}{A^2}-\frac{f_R}{2}\right.\right.\\\nonumber
&\times\left(R-\frac{f}{f_R}\right)\left.+\left(\frac{C'}{C}+\frac{f'_R}{4f_R}-\frac{B'}{B}\left.-\frac{A'}{A}\right)
\frac{f_{R}'}{C^2}\right\}\right]=\left\{\frac{\dot{B}}{B}
\left(\frac{\dot{A}}{A}-\frac{\dot{C}}{C}\right)-\frac{\ddot{C}}{C}\right.
\\\label{14}
&\left.+\frac{\dot{C}\dot{A}}{CA}-\frac{\ddot{B}}{B}\right\}\frac{C^2}{A^2}+\left\{\frac{A'}{A}\left(\frac{B'}{B}-\frac{C'}{C}\right)
+\frac{B''}{B}-\frac{B'C'}{BC}+\frac{A''}{A}\right\}\frac{B^2}{C^2},\\\nonumber
&\frac{\kappa}{f_{R}}\left[C^{2}({P_z}+{\varepsilon})-\frac{s^2}{8\pi
B^4C^2}+\frac{C^2}
{\kappa}\left\{\frac{\ddot{f_R}}{A^2}-\frac{f_{R}'}{C^2}\left(\frac{A'}{A}
+\frac{9f'_R}{4f_R}+\frac{2B'}{B}\right)
-\frac{f_R}{2}\right.\right.\\\nonumber
&\times\left(R-\frac{f}{f_R}\right)\left.\left.+\frac{\dot{f_{R}}}{A^2}+\left(\frac{2\dot{B}}{B}
-\frac{\dot{f_R}}{4f_R}-\frac{\dot{A}}{A}\right)
\right\}\right]=\left\{\left(\frac{2\dot{A}}{A}-\frac{\dot{B}}{B}\right)
\frac{\dot{B}}{B}-\frac{2\ddot{B}}{B}\right\}\frac{C^2}{A^2}\\\label{15}
&+\frac{B'}{B}\left(\frac{B'}{B}+\frac{2A'}{A}\right).
\end{align}
To describe the quantity of matter within the planar system, the
mass function can be evaluated through Taub mass formalism in the
presence of electromagnetic field as \cite{19}
\begin{equation}\label{16}
m(t,z)=\frac{(g)^{\frac{3}{2}}}{2}{R^{~~12}_{12}}+\frac{s^2}{2B}
=\frac{B}{2}\left(\frac{\dot{B}^2}{A^2}-\frac{B'^2}{C^2}\right)+\frac{s^2}{2B}
\end{equation}
which can be written in an alternative way using the fluid velocity
as
\begin{eqnarray}\label{19}
\mathbb{E}\equiv\frac{B'}{C}=\sqrt{U^{2}-\frac{2m(t,r)}{B}+\frac{s^2}{B^2}}.
\end{eqnarray}
Using Eqs.(\ref{12})-(\ref{15}), the temporal and radial variations
of mass function leads to
\begin{align}\nonumber
D_{T}m&=-\frac{\kappa}{2f_{R}}\left\{U\left(\hat{P_z}
-2\pi{E}^2+\frac{{\mathcal{T}_{33}}}{C^2}\right)+\mathbb{E}\left(\hat{q}
-\frac{{\mathcal{T}_{03}}}{AC}\right)\right\}B^2\\\label{17}
&+8\pi^2B^2E(2{\dot{E}}B+3{E}\dot{B}),\\\nonumber
D_{B}m&=\frac{\kappa}{2f_{R}}\left\{\hat{\mu}+\frac{
{\mathcal{T}_{00}}}{A^2}+2\pi{E}^2+\frac{U}{\mathbb{E}}\left(\hat{q}
-\frac{{\mathcal{T}_{03}}}{CA}\right)\right\}B^2+8\pi
B^2{E}B'\\\label{18}&\times(2B{E}'+3{E}B'),
\end{align}
where $U$ is the velocity of the collapsing matter defined by
$U=D_{T}B$ where $(D_{T}=\frac{1}{A} \frac{\partial}{\partial t})$
while $\hat{P_r}=P_r+{\varepsilon},~\hat{q}=q+{\varepsilon},~
\hat{\mu}=\mu+{\varepsilon}$ and
$D_{B}=\frac{1}{B'}\frac{\partial}{\partial r}$ represents radial
derivative operator, respectively. Here E denotes electric field
intensity. It is interesting to indicate that for planar celestial
configuration undergoing collapse, $U$ is chosen to be less than
unity. The link between matter variables and mass function can be
found through integration of Eq.(\ref{18}) with Palatini $f(R)$
background as
\begin{eqnarray}\nonumber
\frac{3m}{B^3}&=&\frac{3\kappa}{2B^3}\int^z_{0}\left[\frac{1}{f_{R}}
\left\{\hat{\mu}+2\pi{E}^2+\frac{{\mathcal{T}_{00}}}{A^2}
+\left(\hat{q}-\frac{{\mathcal{T}_{03}}}{CA}\right)
\frac{U}{\mathbb{E}}\right\}B^2B'\right.\\\label{20}&+&\left.8\pi
EB^2(2BE'+3EB')\right]dz.
\end{eqnarray}
The electric component of the Weyl tensor in terms of unit four
velocity and radial four vector is given as
\begin{equation*}\nonumber
E_{\alpha\beta}=\mathcal{E}\left[\chi_{\alpha}\chi_{\beta}-\frac{1}{3}
(g_{\alpha\beta}+V_\alpha V_\beta)\right],
\end{equation*}
where
\begin{align}\nonumber
\mathcal{E}&=\left[\frac{\ddot{B}}{B}+\left(\frac{\dot{C}}{C}
-\frac{\dot{B}}{B}\right)\left(\frac{\dot{B}}{B}+\frac{
\dot{A}}{A}\right)-\frac{\ddot{C}}{C}\right]\frac{1}{A^{2}}
\\\label{21}
&-\left[\frac{C''}{C}-\left(\frac{C'}{C}+\frac{B'}{B}\right)
\left(\frac{B'}{B}-\frac{A'}{A}\right)-\frac{A''}{A}\right]\frac{1}{C^{2}}
\end{align}
is the scalar encapsulating the effects of spacetime curvature.
Alternatively, using Eqs.(\ref{12}) and (\ref{14})-(\ref{16}), we
get
\begin{align}\label{22}
\frac{3m}{B^3}&=\frac{\kappa}{2f_R}\left(\hat{\mu}-\hat{\Pi}
+\frac{{\mathcal{T}_{00}}}{A^2}-\frac{{\mathcal{T}_{33}}}{C^2}
+\frac{{\mathcal{T}_{11}}}{B^2}+6\pi {E}^2\right)-\mathcal{E},
\end{align}
where $\hat{\Pi}=\hat{P}_z-P_\bot$. The above equation determines
the gravitational contribution of planar geometry due to its fluid
variables, mass function and $f(R)$ extra curvature terms.

\section{Dynamical and Evolution Equations}

In this section, we will establish some scalar functions in the
background of a well-consistent $f(R)$ model. We then make a
correspondence between fluid parameters and Weyl scalar with
Palatini $f(R)$ corrections by constructing modified Ellis
equations. In order to discuss the dynamical properties framed
within modified cosmology, we take the $f(R)$ model as follows
\cite{21a}
\begin{equation}\label{23}
f(R)=R-\mu R_c\tanh\left(\frac{R}{R_c}\right),
\end{equation}
where $\mu$ and $R_c$ are positive constants. The values of these
free parameters are, $R_c\sim H_0^2\sim\frac{8\rho_c}{3m_p^2}\simeq
10^{-84}\textmd{GeV}^2,$ where $R_c$ is roughly of the same order as
the Ricci scalar today, $H_0$ is the present day value of Hubble
constant, and the critical density $\rho_c\simeq
10^{-29}\textmd{gr/cm}^3\sim4.5\times10^{-47}\textmd{GeV}^4$. We
categorize the Riemann tensor in terms of second rank tensors, i.e.,
$X_{\alpha\beta}$ and $Y_{\alpha\beta}$, to devise modified form of
structure scalars as \cite{20}
\begin{equation*}
X_{\alpha\beta}=~^{*}R^{*}_{\alpha\mu\beta\nu}V^{\mu}V^{\nu}=
\frac{1}{2}\eta^{\epsilon\rho}_{~~\alpha\mu}R^{*}_{\epsilon
\rho\beta\nu}V^{\mu}V^{\nu},\quad
Y_{\alpha\beta}=R_{\alpha\mu\beta\nu}V^{\mu}V^{\nu},
\end{equation*}
where left, right and double dual of the Riemann curvature tensor
can be respectively written in a standard form as
\begin{equation*}
^{*}R_{\alpha\beta\gamma\delta}\equiv\frac{1}{2}\eta_{
\alpha\beta\varepsilon\rho}R^{\varepsilon\rho}_{~~\gamma\delta},\quad
R^{*}_{\alpha\beta\gamma\delta}\equiv\frac{1}{2}\eta_{\varepsilon
\rho\gamma\delta}R^{\epsilon\rho}_{~~\alpha\beta},\quad
^{*}R^{*}_{\alpha\beta\gamma\delta}\equiv\frac{1}{2}\eta_{\alpha\beta}
^{~~\varepsilon\rho}R^{*}_{\varepsilon\rho\gamma\delta}.
\end{equation*}
The above tensors can further disintegrate into their trace and
trace-free components as
\begin{align}\label{24}
X_{\alpha\beta}&=\frac{1}{3}X_{T}h_{\alpha\beta}
+X_{TF}\left(\chi_{\alpha}\chi_{\beta}-\frac{1}{3}h_{\alpha\beta}
\right),\\\label{25}
Y_{\alpha\beta}&=\frac{1}{3}Y_{T}h_{\alpha\beta}+Y_{TF}\left(
\chi_{\alpha}\chi_{\beta} -\frac{1}{3}h_{\alpha\beta}\right).
\end{align}
By making use of Eqs.(\ref{12}), (\ref{14}), (\ref{15}) and
(\ref{23})-(\ref{25}), these scalar functions can be written in
terms of fluid variables as
\begin{align}\label{26}
&X_{T}=\frac{\kappa{R_c}(R_c^2+\tilde{R}^2)^{(n+1)}}
{R_c(R_c^2+\tilde{R}^2)^{(n+1)}-2n{\lambda}\tilde{R}R_c^{(2n+2)}}
\left(\hat{\mu}+\frac{\delta_\mu}{A^2}\right),\\\label{27}
&X_{TF}=-\mathcal{E}-\frac{\kappa{R_c}(R_c^2+\tilde{R}^2)^{(n+1)}}
{2[R_c(R_c^2+\tilde{R}^2)^{(n+1)}-2n{\lambda}\tilde{R}R_c^{(2n+2)}]}
\left(\hat{\Pi}-2W{\eta}+\frac{\delta_{P_z}}{C^2}
-\frac{\delta_{P_\bot}}{B^2}\right),\\\label{28}
&Y_{T}=\frac{\kappa{R_c}(R_c^2+\tilde{R}^2)^{(n+1)}}
{2[R_c(R_c^2+\tilde{R}^2)^{(n+1)}-2n{\lambda}\tilde{R}R_c^{(2n+2)}]}
\left(\hat{\mu}+\frac{\delta_\mu}{A^2}+\frac{\delta_{P_z}}{C^2}
+\frac{2\delta_{P_\bot}}{B^2}+3\hat{P_{r}}-2\hat{\Pi}\right),\\\label{29}
&Y_{TF}=\mathcal{E}-\frac{\kappa{R_c}(R_c^2+\tilde{R}^2)^{(n+1)}}
{2[R_c(R_c^2+\tilde{R}^2)^{(n+1)}-2n{\lambda}\tilde{R}R_c^{(2n+2)}]}
\left(\hat{\Pi}-2{\eta}W+\frac{\delta_{P_z}}{C^2}-\frac{\delta_{P_\bot}
}{B^2}\right),
\end{align}
where $\delta_{\mu},~\delta_{P_z}$ and $\delta_{P_\bot}$ are the
corresponding values of dark source components evaluated by taking
into account Eq.(\ref{23}) and given in the Appendix \textbf{A}. We
found that trace part of the second dual of the Riemann tensor has
its dependence on the energy density profile of planar geometry with
some extra curvature terms due to $f(R)$ Palatini gravity while the
remaining scalar functions have their dependence on anisotropic
stress tensor. The conservation of energy and momentum from the
contracted Bianchi identities with ordinary and effective matter
fields
\begin{align}\nonumber
\left(\overset{(D)}
{T^{\alpha\beta}+T^{\alpha\beta}}\right)_{;\beta}=0,\quad
\left(\overset{(D)}
{T^{\alpha\beta}+T^{\alpha\beta}}\right)_{;\beta}=0
\end{align}
yields the couple of equations as follows
\begin{align}\nonumber
&\frac{\dot{\hat{\mu}}}{A}+\frac{\hat{q}'}{C}+\frac{1}{A}\left(\frac{
\dot{C}}{C}+\frac{\dot{f_R}}{2f_R}\right)(\hat{P_z}+\mu)+\frac{
1}{A}(\mu+P_\bot)\left(\frac{2\dot{B}}{B}+\frac{\dot{f_R}}
{f_R}\right)+\frac{\mu\dot{f_R}}{Af_R}\\\label{30}
&+\frac{\hat{q}}{C}\left(\frac{2A'}{A}+\frac{3f'_R}{f_R}+\frac{
2B'}{B}\right)+\frac{D_0(t,r)}{\kappa}+\frac{6\pi{E}^2\dot{F}}{A^2F}=0,\\\nonumber
&\frac{\dot{\hat{q}}}{C}+\frac{\hat{P_z}'}{C}+\frac{1}{C}\left(\frac{
A'}{A}+\frac{f'_R}{2f_R}\right)(\mu+\hat{P_z})+\left(\frac{2B'}
{B}+\frac{f'_R}{f_R}\right)(\hat{P_z}-P_{\bot})\frac{1}{C}
+\frac{\hat{P_z}f'_R}{Cf_R}\\\label{31}
&+\frac{\hat{q}}{A}\left(\frac{2\dot{C}}{C}+\frac{3\dot{f_R}}
{f_R}+\frac{2\dot{B}}{B}\right)+\frac{D_1(t,r)}{\kappa}-\frac{4\pi{E}}{BC^2}
(B{E}'+2{E}B')-4\pi\frac{{E}^2F'}{FC^2}=0,
\end{align}
where the terms $D_0$ and $D_1$ emerge due to Palatini $f(R)$
gravity and are mentioned in Appendix \textbf{A}. Next, we continue
our investigation by constructing couple of differential equations
using the procedure adopted by Ellis \cite{d39}. These equations are
found by using Eqs.(\ref{12})-(\ref{15}), (\ref{17}), (\ref{18}) and
(\ref{23}) and defines a link between matter variables with Palatini
$f(R)$ extra curvature terms and Weyl tensor as
\begin{align}\nonumber
&\left[\mathcal{E}-\frac{\kappa}{2\left[1-\mu
\textmd{sech}^2\left(\frac{R}{R_c}\right)\right]}\left(\hat{\mu}
-\hat{\Pi}+6\pi
E^2+\frac{\delta_\mu}{A^2}-\frac{\delta_{P_z}}{C^2}+\frac{
\delta_{P_\bot}}{B^2}\right)\right]^.\\\nonumber
&=\frac{3\dot{B}}{B}\left[\frac{\kappa}{2\left[1-\mu
\textmd{sech}^2\left(\frac{R}{R_c}\right)\right]}\left(\hat{\mu}+
{P_\bot}+\frac{\delta_\mu}{A^2}+\frac{\delta_{P_\bot}}{B^2}
\right)+4\pi E^2-\mathcal{E}\right]\\\label{32}
&+\frac{3\kappa}{2\left[1-\mu
\textmd{sech}^2\left(\frac{R}{R_c}\right)\right]}\left(\frac{AB'}{BC}\right)
\left(\hat{q}-\frac{\delta_q}{AB}\right)+\frac{24\pi^2E}{B}\left(2B\dot{E}
+3E\dot{B}\right),\\\nonumber
&\left[\mathcal{E}-\frac{\kappa}{2\left[1-\mu
\textmd{sech}^2\left(\frac{R}{R_c}\right)\right]}\left(\hat{\mu}-\hat
{\Pi}+6\pi E^2+\frac{\delta_\mu}{A^2}-\frac{\delta_{P_z}}{C^2}
+\frac{\delta_{P_\bot}}{B^2}\right)\right]_{,1}\\\nonumber
&=-\frac{3{B'}}{B}\left[\frac{\kappa}{2\left[1-\mu
\textmd{sech}^2\left(\frac{R}{R_c}\right)\right]}
\left(\hat{\mu}+2\pi E^2+\frac{\delta_\mu}{A^2}\right)-\frac{3m}
{B^3}\right]\\\label{33}
&-\frac{\kappa}{2\left[1-\mu\textmd{sech}^2\left(\frac{R}{R_c}\right)\right]}
\left(\frac{\dot{B}C}{AB}\right)\left(\hat{q}-\frac{\delta_q}{CA}\right)
+\frac{24\pi^2E}{B}\left(2E'B+3EB'\right),
\end{align}
where $\delta_q$ is mentioned in Appendix \textbf{A}. The limit
$f(R)\rightarrow R$ in the above equations provides GR Ellis
equations.

\section{Irregularities in the Dynamical System}

This section explores some fluid variables that are responsible for
irregularities in the dynamical system having planar symmetry. This
analysis has been carried out from initial homogeneous configuration
of compact body by means some particular choices on matter fields
with extra curvature terms of Palatini $f(R)$ gravity. We will
restrict our analysis on the present day value of cosmological Ricci
scalar, i.e., $R=\tilde{R}$ while dealing with bulky system of
equations. Finally, we will study the case with zero expansion. We
classify our investigation in two scenarios, i.e., dissipative and
non-dissipative systems as follows:

\subsection{Non-radiating Matter}

This section deals with non-dissipative choices of matter fields
like dust, perfect and anisotropic fluid configurations,
respectively, in the Palatini $f(R)$ gravity back ground.

\subsubsection{Dust Fluid}

In this case, we consider $\hat{P}_z=0=P_\bot=\hat{q}$ and $A=1$
indicating geodesic motion of non-dissipative dust cloud. In this
scenario, the two differential equations for Weyl tensor obtained in
(\ref{32}) and (\ref{33}) reduce to
\begin{align}\nonumber
&\left[\mathcal{E}-\frac{\kappa}{2\left[1-\mu
\textmd{sech}^2\left(\frac{\tilde{R}}{R_c}\right)\right]}
\left\{{\mu}+6\pi E^2-\frac{\mu
\tilde{R}}{2}+\frac{\mu}{2}\tanh\left(\frac{\tilde{R}}{R_c}\right)\right.\right.\\\nonumber
&\times\left.\left.\left\{\tilde{R}
\tanh\left(\frac{\tilde{R}}{R_c}\right)-R_c\right\}\right\}\right]^.
=\frac{3\dot{B}}{B}\left[\frac{\kappa}{2\left[1-\mu
\textmd{sech}^2\left(\frac{\tilde{R}}{R_c}\right)\right]}\left(\mu+4\pi
E^2\right)\right]\\\label{34}
&+\frac{24\pi^2E}{B}\left(2\dot{E}B+3E\dot{B}\right),\\\nonumber
&\left[\mathcal{E}-\frac{\kappa}{2\left[1-\mu
\textmd{sech}^2\left(\frac{\tilde{R}}{R_c}\right)\right]}
\left\{{\mu}+6\pi E^2-\frac{\mu
\tilde{R}}{2}+\frac{\mu}{2}\tanh\left(\frac{\tilde{R}}{R_c}\right)\right.\right.\\\nonumber
&\times\left.\left.\left\{\tilde{R}
\tanh\left(\frac{\tilde{R}}{R_c}\right)-R_c\right\}\right\}\right]'
=-\frac{3{B'}}{B}\mathcal{E}-\frac{6\kappa\pi E^2B'}{\left[1-\mu
\textmd{sech}^2\left(\frac{\tilde{R}}{R_c}\right)\right]B}+\frac{24\pi^2E}{B}\\\label{35}
&\times\left(2E'B+3EB'\right).
\end{align}
When $\mu'=0$, Eq.(\ref{35}) leads to
\begin{align}\label{37}
&{\mathcal{E}}'+\frac{3{B'}}{B}\mathcal{E}=\frac{6\pi\kappa
E}{\left[1-\mu
\textmd{sech}^2\left(\frac{\tilde{R}}{R_c}\right)\right]}\left(E'-\frac{EB'}{B}
\right)+\frac{24E\pi^2}{B}\left(2E'B+3EB'\right).
\end{align}
The general solution of the above equation is obtained as
\begin{align}\label{38}
&{\mathcal{E}}=\frac{6\pi}{B^3}\int^z_0\left[\frac{\kappa
EB^3}{\left[1-\mu
\textmd{sech}^2\left(\frac{\tilde{R}}{R_c}\right)\right]}\left(E'-\frac{EB'}{B}\right)+3\pi
EB^2\left(2E'B+3E'B\right)\right]dz.
\end{align}
It is worth noting that Weyl scalar is the only geometric entity
responsible for the irregularities in the energy density depending
upon the electromagnetic profile. In the absence of electromagnetic
field, the Weyl scalar will also vanish showing the importance of
charged fields. By making use of Eqs.(\ref{11}), (\ref{30}) and
(\ref{B3})-(\ref{B6}) in Eq.(\ref{34}), we found
\begin{align}\nonumber
&\dot{\mathcal{E}}+\frac{3\dot{B}}{B}\mathcal{E}=\frac{-\kappa\mu
W_{GR}}{2\left[1-\mu
{\tanh}^2\left(\frac{\tilde{R}}{R_c}\right)\right]}+\frac{6\pi^2\kappa
E}{\left[1-\mu
{\tanh}^2\left(\frac{\tilde{R}}{R_c}\right)\right]}\left(\dot{E}+\frac{E\dot{B}}{B}\right)\\\label{36}
&+\frac{24\pi^2E}{B} \left(2\dot{E}B+3E\dot{B}\right).
\end{align}
The above equation reveal the relationship of Weyl scalar with shear
scalar indicating the shearing motion of dust cloud. It also shows
that the system will be homogeneous if it is shear free as well as
conformally flat within the Palatini framework of $f(R)$ gravity.
Its solution turns out to be
\begin{align}\nonumber
{\mathcal{E}}&=\frac{\kappa}{2B^3\left[1-\mu
\textmd{sech}^2\left(\frac{\tilde{R}}{R_c}\right)\right]}\int_0^t\left[-\mu
W_{GR}+12\pi
E\left(\frac{E\dot{B}}{B}+\dot{E}\right)\right]B^3dt\\\label{39}&+\frac{12\pi^2}{B^3}
\int_0^t\left[EB^3\left(2\dot{E}B+3E\dot{B}\right)\right]dt.
\end{align}
The role of expansion can be brought forward while discussing the
inhomogeneities on the evolution of dust matter in the collapse
scenario. We study the zero expansion case, i.e., $\Theta_P=0$, so
that the above equation becomes
\begin{align}\nonumber
{\mathcal{E}}&=\frac{\kappa}{2B^3\left[1-\mu
\textmd{sech}^2\left(\frac{\tilde{R}}{R_c}\right)\right]}\int_0^t\left[\frac{3\mu\dot{B}}{B}+12\pi
E\left(\frac{E\dot{B}}{B}+\dot{E}\right)\right]B^3dt\\\nonumber&+\frac{12\pi^2}{B^3}
\int_0^t\left[EB^3\left(2\dot{E}B+3E\dot{B}\right)\right]dt.
\end{align}
It shows that the expansion free system will be inhomogeneous due to
the presence of Weyl scalar as it produces tidal forces which made
the object inhomogeneous with the passage of time thus indicating
the importance of time. Moreover, in the absence of tidal forces,
the system will be inhomogeneous due to the presence of
electromagnetic field. Consequently, an expansion free system will
be homogeneous if it is charged free and conformally flat.

\subsubsection{Isotropic Fluid}

In this case, we introduce a bit of complexity in the previous case
by adding the effects of isotropic pressure and determine the
inhomogeneity factors. In this scenario, the Ellis equations
(\ref{32}) and (\ref{33}) turn out to be
\begin{align}\nonumber
&\left[\mathcal{E}-\frac{\kappa}{2\left[1-\mu
\textmd{sech}^2\left(\frac{\tilde{R}}{R_c}\right)\right]}\left[\mu+6\pi
E^2-\frac{\mu
\tilde{R}}{2}+\frac{\mu}{2}\tanh\left(\frac{\tilde{R}}{R_c}\right)\left\{\tilde{R}\tanh\left(
\frac{\tilde{R}}{R_c}\right)\right.\right.\right.\\\nonumber
&\left.\left.\left.-R_c\right\}\right]\right]^.=\frac{3\dot{B}}{B}\left[\frac{\kappa}{2\left[1-\mu
\textmd{sech}^2\left(\frac{\tilde{R}}{R_c}\right)\right]}\left(\mu+P+4\pi
E^2\right)-\mathcal{E}\right]+\frac{24\pi^2E}{B}\\\label{40} &\times
\left(2\dot{E}B+3E\dot{B}\right),\\\nonumber
&\left[\mathcal{E}-\frac{\kappa}{2\left[1-\mu
\textmd{sech}^2\left(\frac{\tilde{R}}{R_c}\right)\right]}\left[\mu+6\pi
E^2-\frac{\mu\tilde{R}}{2}+\frac{\mu}{2}\tanh\left(\frac{\tilde{R}}{R_c}\right)\left\{
\tilde{R}\tanh\left(\frac{\tilde{R}}{R_c}\right)\right.\right.\right.\\\label{41}
&\left.\left.\left.-R_c\right\}\right]\right]'=-\frac{3B'}{B}\left[\mathcal{E}-\frac{2\kappa\pi
E^2}{\left[1-\mu
\textmd{sech}^2\left(\frac{\tilde{R}}{R_c}\right)\right]}\right]+\frac{24\pi^2E}{B}(2E'B+2EB').
\end{align}
We see that the second equation is the same as we have evaluated in
the above case with dust cloud (see Eq.(\ref{34})). Therefore,
indicating the Weyl scalar as the factor responsible of
irregularities in the matter distribution. By making use of
Eqs.(\ref{11}) and (\ref{30}), Eq.(\ref{40}) leads to
\begin{align}\nonumber
\dot{\mathcal{E}}+\frac{3\dot{B}}{B}\mathcal{E}&=\frac{\kappa}{2\left[1-\mu
\textmd{sech}^2\left(\frac{\tilde{R}}{R_c}\right)\right]}\left[-W_{GR}(\mu+P)A+12\pi
E\left(\dot{E}+\frac{E\dot{B}}{B}\right)\right]\\\label{42}
&+\frac{24\pi^2E}{B}(2\dot{E}B+3E\dot{B}),
\end{align}
which on integration turns out to be
\begin{align}\nonumber
{\mathcal{E}}&=\frac{\kappa}{2B^3\left[1-\mu
\textmd{sech}^2\left(\frac{\tilde{R}}{R_c}\right)\right]}\int_0^t\left[-W_{GR}(\mu+P)A+
12\pi
E\left(\dot{E}+\frac{E\dot{B}}{B}\right)\right]B^3dt\\\label{43}&+\frac{24\pi^2}{B^3}\int_0^t\left[
EB^2(2\dot{E}B+3E\dot{B})\right]dt.
\end{align}
This indicates the importance of shear on the evolution of
inhomogeneous matter configuration with isotropic pressure. We
observed that not only shear and pressure, but extra curvature terms
due to $f(R)$ gravity are acting on the system to make it
inhomogeneous as the evolution proceeds. We can also examine the
factors responsible for irregularities over the relativistic system
with zero shear. Moreover, we have already obtained a relation
between expansion and shear scalar, therefore, we can analyze the
those effects when the system is undergoing collapse with zero
expansion. With zero expansion condition, Eq.(\ref{42}) provides
\begin{align}\nonumber
{\mathcal{E}}&=\frac{3\kappa}{2B^3\left[1-\mu
\textmd{sech}^2\left(\frac{\tilde{R}}{R_c}\right)\right]}\int_0^t\left
[\frac{\dot{B}}{B}\left\{(\mu+P)A+4\pi
E^2\right\}+12\pi
E\dot{E}\right]B^3dt\\\label{44}&+\frac{24\pi^2}{B^3}\int_0^t\left[
EB^2(2\dot{E}B+3E\dot{B})\right]dt.
\end{align}
It is seen from the above expression that electromagnetic field have
also crucial role in the expansion free scenario. The Weyl scalar
also plays a key role due to tidal forces making the system more
inhomogeneous with the passage of time.

\subsubsection{Anisotropic Fluid}

This case generalizes the previous one by introducing the complexity
in the form of anisotropic stresses while the dissipative effects
are assumed to be zero, i.e., $\Pi\neq0$ and $\hat{q}=0$. In this
framework, the two equations obtained in (\ref{32}) and (\ref{33})
takes the form as
\begin{align}\nonumber
&\left[\mathcal{E}-\frac{\kappa}{2\left[1-\mu
\textmd{sech}^2\left(\frac{\tilde{R}}{R_c}\right)\right]}\left[
\mu-\Pi+6\pi E^2-\frac{\mu
R}{2}+\frac{\mu}{2}\tanh\left(\frac{R}{R_c}\right)\right.\right.\\\nonumber
&\times\left.\left.
\left\{R\tanh\left(\frac{R}{R_c}\right)-R_c\right\}\right]\right]_{,0}
=\frac{3\dot{B}}{B}\left[\frac{\kappa}{2\left[1-\mu
\textmd{sech}^2\left(\frac{\tilde{R}}{R_c}\right)\right]}
\left\{\mu+P_\bot+4\pi E^2\right\}\right.\\\label{45}
&\left.-\mathcal{E}\right]+\frac{24\pi^2E}{B}(2\dot{E}B+3E\dot{B}),\\\nonumber
&\left[\mathcal{E}-\frac{\kappa}{2\left[1-\mu
\textmd{sech}^2\left(\frac{\tilde{R}}{R_c}\right)\right]}\left[\mu+6\pi
E^2-\Pi+\frac{\mu
R}{2}+\frac{\mu}{2}\tanh\left(\frac{R}{R_c}\right)\right.\right.\\\nonumber
&\left.\left.\left\{R\tanh\left(\frac{R}{R_c}\right)-R_c\right\}\right]\right]'
-\frac{3B'}{B}\left[\mathcal{E}+\frac{\kappa}{2\left[1-\mu
\textmd{sech}^2\left(\frac{\tilde{R}}{R_c}\right)\right]}\left[\Pi-4\pi
E^2\right]\right]\\\label{46}&+\frac{24\pi^2E}{B}\left(2E'B+3EB'\right).
\end{align}
We can found the following couple of equations by using
Eqs.(\ref{11}) and (\ref{30}) in (\ref{45}) and (\ref{46}) with some
computation as
\begin{align}\nonumber
&\left[\mathcal{E}+\frac{\kappa}{2\left[1-\mu
\textmd{sech}^2\left(\frac{\tilde{R}}{R_c}\right)\right]}(\Pi-4\pi
E^2)\right]^.+\frac{3\dot{B}}{B}\left[\frac{\kappa}{2\left[1-\mu
\textmd{sech}^2\left(\frac{\tilde{R}}{R_c}\right)\right]}\right.(\Pi\\\nonumber
&-4\pi E^2)\left.+\mathcal{E}\right]=\frac{-\kappa}{2\left[1-\mu
\textmd{sech}^2\left(\frac{\tilde{R}}{R_c}\right)\right]}(\mu+P_z)AW_{GR}
+\frac{3\kappa\pi\dot{B}E^2}{\left[1-\mu
\textmd{sech}^2\left(\frac{\tilde{R}}{R_c}\right)\right]B}\\\nonumber
&+\frac{\kappa\pi\dot{B}}{\left[1-\mu
\textmd{sech}^2\left(\frac{\tilde{R}}{R_c}\right)\right]B}+\frac{2\kappa\pi\dot{E}E}{\left[1-\mu
\textmd{sech}^2\left(\frac{\tilde{R}}{R_c}\right)\right]},\\\nonumber
&\left[\mathcal{E}-\frac{\kappa}{2\left[1-\mu
\textmd{sech}^2\left(\frac{\tilde{R}}{R_c}\right)\right]}(\mu+\Pi-2\pi
E^2)\right]'=-\frac{3B'}{B}\left[\frac{\kappa}{2\left[1-\mu
\textmd{sech}^2\left(\frac{\tilde{R}}{R_c}\right)\right]}\right.\\\nonumber
&\times\left.(\Pi-4\pi
E^2)+\mathcal{E}\right]+\frac{24\pi^2E}{B}(2E'B+3EB').
\end{align}
Using the trace free part of the second dual of Riemann tensor as
obtained in Eq.(\ref{27}), we found
\begin{align}\nonumber
\dot{X}_{TF}+\frac{3X_{TF}\dot{B}}{B}&=\frac{\kappa}{\left[1-\mu
\textmd{sech}^2\left(\frac{\tilde{\tilde{R}}}{R_c}\right)\right]}\left\{\frac{AW_{GR}}
{2}(\mu+P_z)-\frac{\Pi\dot{B}}{B}\right.\\\nonumber
&\left.+\pi
E\left(\frac{3E\dot{B}}{B}+2\dot{E}\right)\right\},\\\nonumber
X'_{TF}+\frac{3X_{TF}{B'}}{B}&=\frac{\kappa}{2\left[1-\mu
\textmd{sech}^2\left(\frac{\tilde{R}}{R_c}\right)\right]}\left[\mu'+4\pi
EE'\right]+\frac{24\pi^2E}{B}(2E'B+3EB').
\end{align}
The solution of the the above couple of differential equations turn
out to be
\begin{align}\nonumber
X_{TF}&=-\frac{\kappa}{\left[1-\mu
\textmd{sech}^2\left(\frac{\tilde{R}}{R_c}\right)\right]}\int_0^t\left[\Pi\dot{B}-\frac{AW_{GR}}{2}(\mu+P_z)B-\pi
EB\left(\frac{3E\dot{B}}{B}\right.\right.\\\label{47}
&\left.\left.+2\dot{E}\right)\right]B^2dt,\\\nonumber
X_{TF}&=\frac{\kappa}{2\left[1-\mu
\textmd{sech}^2\left(\frac{\tilde{R}}{R_c}\right)\right]}\int_0^z(\mu'+4\pi
EE')B^3dz+\frac{24\pi^2}{B^3}\int_0^zEB^2\\\label{48}&\times(2E'B+3EB')dz.
\end{align}
Equation (\ref{47}) shows a relation of one of the scalar function,
from the splitting of the Riemann tensor, with anisotropic pressure
and shear scalar. It indicates the importance of these material
variables with planar geometry in the discussion of irregular energy
distribution. Now, the factor, that controls inhomogeneities over
the compact system, is the trace free part of the double dual of the
Riamann tensor, which is obtained through the orthogonal splitting
of the Riemann tensor in the framework of Palatini $f(R)$ gravity as
seen from Eq.(\ref{48}). It is well known that these scalar
functions play a crucial role in the structure formation of the
universe. Also, the solution of the field equations in the static
case can be written in the form of these scalar functions. We found
that in the absence of electromagnetic field, $X_{TF}$ is the factor
describing the irregularities in the star configuration.
Consequently, if $X_{TF}=0$ then the matter distribution in the
charged free system will be homogeneous and vice versa. Next, we
discuss the case of collapsing matter with zero expansion in the
presence of anisotropic pressure. In this scenario, the solution of
Eq.(\ref{45}) becomes
\begin{align}\nonumber
X_{TF}&=-\frac{\kappa}{\left[1-\mu
\textmd{sech}^2\left(\frac{\tilde{R}}{R_c}\right)\right]}\int_0^t\left[\Pi\dot{B}+\frac{3A}{2}(\mu+P_z)\dot{B}-\pi
EB\right.\\\label{49}
&\times\left.\left(\frac{3E\dot{B}}{B}+2\dot{E}\right)\right]B^2dt,
\end{align}
which provides a link of the structure scalar with energy density
and pressure anisotropy in the arrow of time with extra curvature
terms due to Palatini $f(R)$ gravity. Since we know that in the
expansion free system, the center is surrounded by another spacetime
appropriately matched with the rest of the system.

\subsection{Radiating Dust Fluid}

This section explores the inhomogeneity factors with dissipation in
both diffusion and free streaming limit, but with a particular case
of charged dust cloud. For this purpose, we take $P_z=0=P_\bot$ in
the matter field and the motion is considered to be geodesic by
assuming $A=1$ in the geometric part, which is well justified on the
basis of some theoretical advances made in the discussion of
inhomogeneous matter distribution. In this framework, Eqs.(\ref{32})
and (\ref{33}) yield
\begin{align}\nonumber
&\left[\mathcal{E}-\frac{\kappa}{2\left[1-\mu
\textmd{sech}^2\left(\frac{\tilde{R}}{R_c}\right)\right]}\left\{\bar{\mu}+6\pi
E^2-\frac{\mu\tilde{R}}{2}+\frac{\mu}{2}\tanh\left(\frac{\tilde{R}}{R_c}\right)\right.\right.\\\nonumber
&\times\left.\left. \left[
\tilde{R}\tanh\left(\frac{\tilde{R}}{R_c}\right)-R_c\right]\right\}\right]^.
=\frac{3\dot{B}}{B}\left[\frac{\kappa}{2\left[1-\mu
\textmd{sech}^2\left(\frac{\tilde{R}}{R_c}\right)\right]}(\tilde{\mu}+4\pi
E^2)-\mathcal{E}\right]\\\label{50}&+\frac{3\kappa}{2\left[1-\mu
\textmd{sech}^2\left(\frac{\tilde{R}}{R_c}\right)\right]}\frac{qAB'}{BC}
+\frac{24\pi^2E}{B}(2\dot{E}B+3E\dot{B}),\\\nonumber
&\left[\mathcal{E}-\frac{\kappa}{2\left[1-\mu
\textmd{sech}^2\left(\frac{\tilde{R}}{R_c}\right)\right]}\left\{\bar{\mu}+6\pi
E^2-\frac{\mu\tilde{R}}{2}+\frac{\mu}{2}\tanh\left(\frac{\tilde{R}}{R_c}\right)\right.\right.\\\nonumber
&\times\left.\left.\left[
\tilde{R}\tanh\left(\frac{\tilde{R}}{R_c}\right)-R_c\right]\right\}\right]'
=-\frac{3B'}{B}\left[\mathcal{E}+\frac{2\kappa\pi
E^2}{\left[1-\mu
\textmd{sech}^2\left(\frac{\tilde{R}}{R_c}\right)\right]}\right]\\\label{51}
&-\frac{3\kappa\dot{B}C\bar{q}}{2AB'\left[1-\mu
\textmd{sech}^2\left(\frac{\tilde{R}}{R_c}\right)\right]}+\frac{24\pi^2E}{B}(2E'B+3EB').
\end{align}
Consider
\begin{align}\label{52}
&\Psi\equiv\mathcal{E}+\frac{1}{B^3}\int_0^z\frac{3\kappa
B^2C\dot{B}\bar{q}}{2\left[1-\mu
\textmd{sech}^2\left(\frac{\tilde{R}}{R_c}\right)\right]}dz.
\end{align}
If we consider the matter distribution to be homogeneous i.e.,
$\mu'=0$, then from Eq.(\ref{51}) we obtain the following expression
\begin{align}\nonumber
\Psi&=\frac{1}{B^3}\int_0^z\left[6\pi
EB^3E'\left\{\frac{\kappa}{\left[1-\mu
\textmd{sech}^2\left(\frac{\tilde{R}}{R_c}\right)\right]}+8\pi\right\}\right.\\\label{52z}
&+\left.6\pi E^2B^2B'\left\{12\pi-\frac{\kappa}{\left[1-\mu
\textmd{sech}^2\left(\frac{\tilde{R}}{R_c}\right)\right]}\right\}\right]dz,
\end{align}
which should be vanishing for the homogeneous fluid distribution
over planar geometry. Consequently, for homogeneous universe with
planar topology, one should have $\Psi=0\Leftrightarrow \mu'=0$ with
dissipative charged dust cloud. The evolution equation for $\Psi$
can also be evaluated using Eqs.(\ref{11}) and (\ref{30}) in
Eq.(\ref{50}) as
\begin{align}\nonumber
&\dot{\Psi}-\frac{\dot{\Omega}}{B^3}=\frac{\kappa}{2\left[1-\mu
\textmd{sech}^2\left(\frac{\tilde{R}}{R_c}\right)\right]}\left[-\tilde{\mu}W_{GR}-\frac{\tilde{q}}{C}
+\frac{\tilde{q}B'}{BC}\right]-\frac{3\dot{B}\Psi}{B}\\\label{53}
&+6\pi EE'\left(\frac{\kappa}{\left[1-\mu
\textmd{sech}^2\left(\frac{\tilde{R}}{R_c}\right)\right]}+8\pi\right)+\frac{6\pi
E^2\dot{B}}{B}\left(12\pi+\frac{\kappa}{\left[1-\mu
\textmd{sech}^2\left(\frac{\tilde{R}}{R_c}\right)\right]}\right).
\end{align}
whose solution leads to
\begin{align}\nonumber
&\Psi=\frac{1}{B^3}\int_0^t\left[\dot{\Omega}+\frac{\kappa}{2\left[1-\mu
\textmd{sech}^2\left(\frac{\tilde{R}}{R_c}\right)\right]}\left\{-\tilde{
\mu}W_{GR}B-\frac{\tilde{q}B}{C}+\frac{\tilde{q}B'}{C}\right\}+6\pi
EB^3\dot{E}\right.\\\label{54}
&\times\left.\left(\frac{\kappa}{\left[1-\mu
\textmd{sech}^2\left(\frac{\tilde{R}}{R_c}\right)\right]}+8\pi\right)+6\pi
E^2B^2\dot{B}\left(12\pi+\frac{\kappa}{\left[1-\mu
\textmd{sech}^2\left(\frac{\tilde{R}}{R_c}\right)\right]}\right)\right]dt.
\end{align}
This indicates the importance of fluid parameters as the
inhomogeneity factor is related to the matter variables,
particularly heat flux, as well as kinematical quantities of the
system. Since we already found a relation in which the shear scalar
is related to expansion scalar. Thus, for the shear free case, we
can obtain
\begin{align}\nonumber
\frac{\dot{B}}{B}=\frac{A\Theta_p}{3}-\frac{2}{3}\frac{\dot{F}}{F}.
\end{align}
Using the above equation in Eq.(\ref{31}), we found
\begin{align}\nonumber
\dot{q}&+\left[\frac{2\mu
R'}{R_c}sech^2\left(\frac{R}{R_c}\right)\tanh\left(\frac{R}{R_c}\right)\left[1-\mu
\textmd{sech}^2\left(\frac{{R}}{R_c}\right)\right]^{-1}\right]\left(\frac{\mu}{2C}-\frac{4\pi
E^2}{C^2}\right)\\\nonumber &-\frac{4\pi
E}{BC^2}(BE'+2EB')+\frac{2q}{3}\left[2\Theta_p+\frac{\mu\dot{R}}{R_c}
\textmd{sech}^2\left(\frac{R}{R_c}\right)\tanh\left(\frac{R}{R_c}\right)\right.\\\label{56}
&\times\left.\left[1-\mu
\textmd{sech}^2\left(\frac{{R}}{R_c}\right)\right]^{-1}\right]+D_3=0.
\end{align}

Next, the transportation of heat in the system can be analyzed
through a casual radiating theory defined by Muller and Israel as a
second order thermodynamical theory in diffusion approximation as
follows
\begin{align}\nonumber
\tau\dot{q}&=-\frac{1}{2}\xi{q}K^2\left(\frac{\tau}{{\xi}K^2}\right)_{,0}
-\frac{\xi}{B}(AK)'-qA-\frac{1}{2}{\tau}qA\Theta,
\end{align}
whose independent component yield
\begin{align}\label{57}
\dot{q}=-\frac{q}{\tau}-\frac{\kappa}{C\tau}T'.
\end{align}
Substituting the value of $\dot{q}$ from Eq.(\ref{56}) in the above
equation, we obtain
\begin{align}\nonumber
q&=\left[\frac{-4\pi E\tau}{r^2B^3}(BE'+2EB')-\frac{\kappa
T'}{rB}+D_{3s}+\left[\frac{2\mu
R'}{R_c}\textmd{sech}^2\left(\frac{R}{R_c}\right)\tanh\left(\frac{R}{R_c}\right)\right.\right.\\\nonumber
&\times\left.\left.\left[1-\mu
\textmd{sech}^2\left(\frac{{R}}{R_c}\right)\right]^{-1}\right]\left(
\frac{\mu\tau}{2rB}-\frac{4\pi
E^2\tau}{r^2B^2}\right)\right]\left[1-\frac{4\Theta_p\tau}{3}\right].
\end{align}
One can identify the relaxation effects by inserting this value in
the evolution equation for the inhomogeneity factor in this case,
i.e., $\Psi$ as obtained in Eq.(\ref{54}). Consequently, the effects
of electromagnetic field with the relaxation time can also be
analyzed.

\section{Discussion}

In this paper, we have investigated some inhomogeneity factors for
self-gravitating plane symmetric model. We have done this analysis
by taking anisotropic matter distribution in the presence of
electromagnetic field. The particular attention has been given to
examine the role of dark source terms coming from the modification
of gravitational field. The modification includes higher order
curvature terms explicitly due to Palatini $f(R)$ corrections in the
field equations. In order to continue our analysis systematically,
first of all we have explored the Palatini $f(R)$-Maxwell field
equations for our compact object and define the mass function using
Taub's mass formalism. An expression for the Weyl scalar has been
disclosed in terms of matter variables and higher curvature
ingredients due to modified gravity.

A set of scalar functions have been evaluated using the splitting of
Riemann curvature tensor with comoving coordinate system to discuss
the irregularities in energy density. These scalar functions are
named as structure scalars whose physical significance have been
analyzed in the literature previously. Also, it is established that
these scalars are used to write down solutions of field equations
with static background metric. We have related these scalars in
terms of material variables and dark source terms using the field
equations and a cosmological $f(R)$ model. Moreover, a couple of
equations describing the conservation of energy-momentum in space
have been explored. The evolution equations are also investigated
using the procedure adopted by Ellis \cite{22}. We have found some
factors responsible for inhomogeneities in the matter configuration
with some particular cases of fluid distribution.

A particular attention is given to examine the role of
electromagnetic field in this framework. It is usually considered in
the study of relativistic astrophysics that compact objects have no
sufficient internal electric fields. It is still vivid that stars
can have total net charge or large internal electric fields.
However, it is well established that angular momentum plays the role
of electric charge in rotating collapsing stars. In the present
study, we have dropped some light on more realistic astrophysical
scenario, i.e., the inhomogeneities/irregularities in the universe
model. The galaxy distribution is observed to be inhomogeneous at
small scales while, according  to  theoretical  models, it is
expected  to become spatially  homogeneous  for $r>\lambda_0
\approx10 Mpch^{-1}$ \cite{21}.

On the basis of the results we have obtained, it is clear that the
system becomes inhomogeneous as the evolution proceeds with time
indicating a crucial role of gravitational arrow of time. In the
non-radiating dust cloud case, we have found that the system will be
homogeneous in the absence of electromagnetic field as well as tidal
forces, which are due to the presence of Weyl scalar. It shows that
the Weyl tensor and presence of charge make the distribution of
matter more inhomogeneous during evolution of the universe. With the
inclusion of isotropic pressure in matter configuration, the Weyl
tensor and electric charge behave similarly as in the dust case. In
the presence of anisotropic pressure effects in the matter, we have
found a particular factor, known as the trace free component of dual
of the Riemann tensor, responsible for the irregularities in the
planar system. In the radiating dust cloud case, we have found that
the system will be homogeneous if the factor $\Psi$ given in
Eq.(\ref{54}) vanishes otherwise it will make our geometric model
more inhomogeneous with evolution in time.

All of our results reduces to charge-free case \cite{z1} under the
limit $s=0$ while our results support the analysis made by \cite{18}
under the limit $f(R)=R$

\section*{Acknowledgments}

This work was partially supported by University of the Punjab,
Lahore-Pakistan through research project in the fiscal year
2015-2016 (M.Z.B.).

\vspace{0.5cm}

\renewcommand{\theequation}{B\arabic{equation}}
\setcounter{equation}{0}
\section*{Appendix A}

The higher curvature terms $D_0$ and $D_1$ of Eqs.(\ref{30}) and
(\ref{31}) are given as
\begin{align}\nonumber
D_0&=\left.\frac{(-1)}{A^2}\left\{\left(\frac{f}{R}-f_R\right)
\frac{R}{2}-\frac{f''_R}{C^2}+\frac{\dot{f_R}}{A^2}\left(\frac{
\dot{C}}{C}+\frac{9\dot{f_R}}{4f_R}+\frac{2\dot{B}}{B}\right)
-\frac{f'_R}{C^2}\left(\frac{C'}{C}\right.\right.\right.\\\nonumber
&\left.\left.\left.+\frac{f'_R}{4f_R}-\frac{2B'}{B}\right)
\right\}_{,0}+\frac{\dot{f_R}}{f_RA}\left\{\frac{3\ddot{f_R}}
{2A^2}-\frac{R}{2}\left(f_R-\frac{f}{R}\right)+\frac{3f''_R}
{2C^2}-\frac{\dot{f_R}}{A^2}\left(\frac{3\dot{C}}{2C}\right.
\right.\right.\\\nonumber
&\left.\left.\left.+\frac{3\dot{A}}{2A}+\frac{5\dot{B}}{B}
+\frac{6\dot{f_R}}{f_R}\right)-\frac{f'_R}{C^2}\left(\frac{3A'}
{2A}+\frac{3C'}{2C}-\frac{3B'}{B}+\frac{3f'_R}{2f_R}\right)
\right\}+\frac{\dot{C}}{AC}\left\{\frac{f''_R}{C^2}\right.\right.\\\nonumber
&\left.\left.+\frac{\ddot{f_R}}{A^2}-\frac{\dot{f_R}}{A^2}
\left(\frac{5\dot{f_R}}{2f_R}+\frac{\dot{A}}{A}+\frac{4\dot{B}}
{B}+\frac{\dot{C}}{C}\right)-\frac{f'_R}{C^2}\left(\frac{A'}{A}
+\frac{5f'_R}{2f_R}+\frac{C'}{C}\right)\right\}\right.\\\nonumber
&\left.+\frac{(-1)}{C^2A}\left(\dot{f'_R}-\frac{5}{2}\frac{\dot{
f_R}f'_R}{f_R}-\frac{A'}{A}\dot{f_R}-\frac{\dot{C}}{C}f'_R\right)
\left(\frac{3A'}{A}+\frac{C'}{C}+\frac{3f'_R}{f_R}+\frac{2B'}{B}
\right)\right]\\\label{B1}
&+\left.{A}\left[\frac{1}{A^2C^2}\left\{\dot{f'_R}
-\frac{A'}{A}\dot{f_R}-\frac{\dot{C}}{C}f'_R-\frac{5}{2}\frac{
\dot{f_R}f'_R}{f_R}\right\}\right]_{,1}\right.,\\\nonumber
D_1&=\left.C\left\{\frac{-1}{(CA)^2}\left(\dot{f'_R}-\frac{5
\dot{f_R}f'_R}{2f_R}-\frac{A'}{A}\dot{f_R}-\frac{\dot{C}}{C}
f'_R\right)\right\}_{,0}+\frac{1}{C}\left\{\frac{\ddot{f_R}}
{A^2}-\frac{R}{2}\right.\right.\\\nonumber
&\left.\left.\times\left(f_R-\frac{f}{R}\right)-\frac{\dot{f_R}}
{A^2}\left(\frac{\dot{A}}{A}+\frac{\dot{f_R}}{4f_R}+\frac{2
\dot{B}}{B}\right)-\frac{f'_R}{C^2}\left(\frac{2B'}{B}+\frac{
9f'_R}{4f_R}+\frac{A'}{A}\right)\right\}_{,1}\right.\\\nonumber
&+\left.\frac{A'}{CA}\left\{\frac{f''_R}{C^2}+\frac{\ddot{f_R}}
{A^2}-\frac{\dot{f_R}}{A^2}\left(\frac{5\dot{f_R}}{2f_R}+\frac{
\dot{A}}{A}+\frac{4\dot{B}}{B}+\frac{\dot{C}}{C}\right)-\frac{f'_R}
{C^2}\left(\frac{5f'_R}{2f_R}+\frac{A'}{A}\right.\right.\right.\\\nonumber
&\left.\left.\left.+\frac{C'}{C}\right)\right\}+\frac{f'_R}{f_RC}
\left\{\left(\frac{f}{R}-f_R\right)\frac{R}{2}+\frac{3\ddot{f_R}}
{2A^2}+\frac{3f_R''}{2C^2}-\frac{3\dot{f_R}}{2A^2}\left(\frac{
\dot{A}}{A}+\frac{\dot{C}}{C}+\frac{\dot{f_R}}{f_R}\right.\right.\right.\\\nonumber
&\left.\left.\left.+\frac{14\dot{B}}{3B}\right)-\frac{f'_R}{C^2}
\left(\frac{3B'}{B}+\frac{3A'}{2A}+\frac{3C'}{2C}+\frac{6f'_R}
{f_R}\right)\right\}+\frac{2B'}{CB}\left\{\frac{f''_R}{C^2}
-\frac{\dot{f_R}}{A^2}\right.\right.\\\nonumber
&\left.\left.\times\left(\frac{\dot{3B}}{B}+\frac{\dot{C}}{C}
\right)-\frac{f'_R}{C^2}\left(\frac{B'}{B}+\frac{5f'_R}{2f_R}
+\frac{C'}{C}\right)\right\}+\frac{(-1)}{CA^2}\left(-\frac{A'}
{A}\dot{f_R}+\dot{f'_R}\right.\right.\\\label{B2}
&\left.\left.-\frac{5\dot{f_R}f'_R}{2f_R}-\frac{\dot{C}}{C}f'_R\right)
\left(\frac{\dot{A}}{A}+\frac{3\dot{C}}{C}+\frac{3\dot{f_R}}{f_R}
\right)\right.
\end{align}

The quantities $\delta_\mu,~\delta_{P_z},~\delta_{P_\bot}$ and
$\delta_q$ are
\begin{align}\nonumber
&\delta_\mu=\frac{-A^2}{\kappa}\left[\frac{2\mu
R'}{C^2R_c}\textmd{sech}^2\left(\frac{R}{R_c}\right)\tanh\left(\frac{R}{R_c}\right)\left[
\frac{C'}{C}-\frac{2B'}{B}+\frac{\mu
R'}{2R_c}\textmd{sech}^2\left(\frac{R}{R_c}\right)\right.\right.\\\nonumber
&\times\tanh\left(\frac{R}{R_c}\right)\left. \left\{
1-\mu\textmd{sech}^2\left(\frac{R}{R_c}\right)\right\}^{-1}\right]+\frac{\mu
R}{2}-\frac{\mu}{2}\tanh\left(\frac{R}{R_c}\right)\left\{R\tanh\left(\frac{R}{R_c}\right)\right.\\\nonumber
&\left.-R_c\right\}-\frac{\left[1-\mu\textmd{sech}^2\left(\frac{R}{R_c}\right)
\right]}{C^2}+\left[\frac{\dot{C}}{C}+\frac{2\dot{B}}{B}
+\frac{9\mu
R'}{2R_c}\textmd{sech}^2\left(\frac{R}{R_c}\right)\tanh\left(\frac{R}{R_c}\right)\right.\\\label{B3}
&\left.\left.\times\left\{
1-\mu\textmd{sech}^2\left(\frac{R}{R_c}\right)\right\}^{-1}
\right]\left[\frac{2\mu
\dot{R}}{R_c}\textmd{sech}^2\left(\frac{R}{R_c}\right)\tanh\left(\frac{R}{R_c}\right)\right]\right],\\\nonumber
&\delta_q=-\frac{1}{\kappa}\left[\left[\frac{2\mu
\dot{R}}{R_c}\textmd{sech}^2\left(\frac{R}{R_c}\right)\tanh\left(\frac{R}{R_c}\right)\left\{
1-\mu\textmd{sech}^2\left(\frac{R}{R_c}\right)\right\}^{-1}\right]'-\frac{10\mu^2R'\dot{R}}{R_c^2}\right.\\\nonumber
&\times\textmd{sech}^4
\left(\frac{R}{R_c}\right)\tanh^2\left(\frac{R}{R_c}\right)\left\{
1-\mu\textmd{sech}^2\left(\frac{R}{R_c}\right)\right\}^{-1}-\frac{2\mu
R'\dot{C}}{CR_c}\textmd{sech}^2\left(\frac{R}{R_c}\right)\\\label{B4}
&\times\left. \tanh\left(\frac{R}{R_c}\right)\left\{
1-\mu\textmd{sech}^2\left(\frac{R}{R_c}\right)\right\}^{-1}-\frac{2\mu\dot{R}A'}{AR_c}\textmd{sech}^2
\left(\frac{R}{R_c}\right)\tanh\left(\frac{R}{R_c}\right)\right.\\\nonumber
&\times \left.\left\{
1-\mu\textmd{sech}^2\left(\frac{R}{R_c}\right)\right\}^{-1}\right],\\\nonumber
&\delta_{P_\bot}=\frac{B^2}{\kappa}\left[\left[\frac{\dot{B}}{B}-\frac{\dot{A}}{A}+\frac{\dot{C}}{C}
-\frac{\mu
\dot{R}}{2R_c}\textmd{sech}^2\left(\frac{R}{R_c}\right)\tanh\left(\frac{R}{R_c}\right)\left\{
1-\mu\textmd{sech}^2\left(\frac{R}{R_c}\right)\right\}^{-1}\right]\right.\\\nonumber
&\times\left.\frac{2\mu
\dot{R}}{A^2R_c}\textmd{sech}^2\left(\frac{R}{R_c}\right)\tanh\left(\frac{R}{R_c}\right)\left\{
1-\mu\textmd{sech}^2\left(\frac{R}{R_c}\right)\right\}^{-1}+\frac{\left[1-\mu\textmd{sech}^2\left(
\frac{R}{R_c}\right)\right]^{..}}{A^2}\right.\\\nonumber
&-\left.\frac{\left[1-\mu\textmd{sech}^2\left(\frac{R}{R_c}\right)\right]''}{A^2}+\frac{\mu
R}{2}-\frac{\mu}{2}\tanh\left(\frac{R}{R_c}\right)\left\{R\tanh\left(\frac{R}{R_c}\right)-R_c\right\}\right.\\\nonumber
&+\left.\left[\frac{C'}{C}-\frac{B'}{B}-\frac{A'}{A}+\frac{\mu
R'}{2R_c}\textmd{sech}^2\left(\frac{R}{R_c}\right)\tanh\left(\frac{R}{R_c}\right)\left\{
1-\mu\textmd{sech}^2\left(\frac{R}{R_c}\right)\right\}^{-1}\right]\right.\\\label{B5}
&\times\left.\frac{2\mu
R'}{C^2R_c}\textmd{sech}^2\left(\frac{R}{R_c}\right)\tanh\left(\frac{R}{R_c}\right)\right],\\\nonumber
&\delta_{P_z}=\frac{C^2}{\kappa}\left[\frac{\mu
R}{2}-\frac{\mu}{2}\tanh\left(\frac{R}{R_c}\right)\left\{R\tanh\left(\frac{R}
{R_c}\right)-R_c\right\}+\frac{\left[1-\mu\textmd{sech}^2\left(
\frac{R}{R_c}\right)\right]^{..}}{A^2}\right.\\\nonumber
&-\left.\left[\frac{A'}{A}+\frac{2B'}{B}+\frac{9\mu
R'}{2R_c}\textmd{sech}^2\left(\frac{R}{R_c}\right)\tanh\left(\frac{R}{R_c}\right)\left\{
1-\mu\textmd{sech}^2\left(\frac{R}{R_c}\right)\right\}^{-1}\right]\right.\\\nonumber
&\times\left.\frac{2\mu
R'}{C^2R_c}\textmd{sech}^2\left(\frac{R}{R_c}\right)\tanh\left(\frac{R}{R_c}\right)+\frac{2\mu
\dot{R}}{A^2R_c}\textmd{sech}^2\left(\frac{R}{R_c}\right)\tanh\left(\frac{R}{R_c}\right)\right.\\\label{B6}
&\times\left.\frac{A'}{A}+\frac{2B'}{B}+\frac{9\mu
R'}{2R_c}\textmd{sech}^2\left(\frac{R}{R_c}\right)\tanh\left(\frac{R}{R_c}\right)\right].
\end{align}


\vspace{0.25cm}

\begin{thebibliography}{40}

\bibitem{z2} Sahni, V. and Starobinsky, A.: Int. J. Mod. Phys. D
\textbf{09}(2000)373; Carroll, S.M.: Living Rev. Relativity
\textbf{4}(2001)1; Padmanabhan, T.: Phys. Rep.
\textbf{380}(2003)235; Riess A. G. et al.: Astrophys. J.
\textbf{659}(2007)98.

\bibitem{z3} Capozziello S. and Laurentis M. D.: Phys. Rep.,
\textbf{509}(2011)167; Nojiri, S. and Odintsov, S.D.: Phys. Rep.
\textbf{505}(2011)59.

\bibitem{z4} Felice A. D. and Tsujikawa S.: Living Rev. Relativity \textbf{13}(2010)3;
Capozziello S., Laurentis M. D. and Faraoni V.: Open Astron. J.
\textbf{3}(2010)49.

\bibitem{1} Kainulainen, K., Reijonen, V. and Sunhede, D.: Phys. Rev. D
\textbf{76}(2007)043503.

\bibitem{2} Nojiri, S. and Odintsov, S.D.: Int. J. Geom. Methods Mod. Phys.
\textbf{04}(2007)115.

\bibitem{3} Guo, J. and Joshi, P.S.: arXiv:1511.06161v1.

\bibitem{4} Nguyen, P.H. and Pedraza, J.F.: Phys. Rev. D
\textbf{88}(2013)064020.

\bibitem{5} Leon, G., Saridakis, E.N.: Class. Quantum Grav. \textbf{28}(2011)065008.

\bibitem{6} Cosenza, M., Herrera, L., Esculpi, M. and Witten, L.: Phys. Rev. D \textbf{25}(1982)2527.

\bibitem{7} Maartens, R., Sahni, V. and Saini, T.D.: Phys. Rev. D
\textbf{63}(2001)063509.

\bibitem{8} Weber, F.: \textit{Pulsars as Astrophysical Observatories for
Nuclear and Particle Physics,} (IOP Publishing, 1999); Martinez,
A.P., Rojas, H.P. and Cuesta, H.J.M.: Eur. Phys. J. C
\textbf{29}(2003)111.

\bibitem{10} Ghezzi, C.R.: Phys. Rev. D \textbf{72}(2005)104017.

\bibitem{11} Varela, V., Rahaman, F., Ray, S., Chakraborty, K. and Kalam, M.: Phys. Rev. D
\textbf{82}(2010)044052.

\bibitem{z5} Sharif, M. and Yousaf, Z.: Phys. Rev. D \textbf{88}(2013)024020; Astropart.
Phys. \textbf{56}(2014)19; Astrophys. Space Sci.
\textbf{352}(2014)321; ibid. \textbf{354}(2014)431; Mon. Not. R.
Astron. Soc. \textbf{440}(2014)3479; J. Cosmol. Astropart. Phys.
\textbf{06}(2014)019.

\bibitem{znew1} Yousaf, Z., Bamba, K. and Bhatti, M.Z.: Phys. Rev. D \textbf{93}(2016)064059;
Yousaf, Z. and Bhatti, M.Z.: Mon. Not. R. Astron. Soc.
\textbf{458}(2016)1785.

\bibitem{12} Penrose, R. and Hawking, S.W.:
\textit{General Relativity, An Einstein Centenary Survey} (Cambridge
University Press, 1979).

\bibitem{13} Herrera, L., Di Prisco, A., Hern\'{a}ndez-Pastora, J.L. and Santos,
N.O.: Phys. Lett. A \textbf{237}(1998)113.

\bibitem{ya29} Herrera,
L., Di Prisco, A., Martin, J., Ospino, J., Santos, N.O. and
Troconis, O.: Phys. Rev. D \textbf{69}(2004)084026; Herrera, L.,
Ospino, J., Di Prisco, A., Ospino, J. and Carot, J.: Phys. Rev. D
\textbf{82}(2010)024021.

\bibitem{ya30} Herrera, L., Di Prisco, A. and  Ib\'{a}\~{n}ez, J.: Phys. Rev. D \textbf{84}(2011)107501.

\bibitem{14} Cai, Y., Zhao, W. and Zhang, Y.: Phys. Rev. D
\textbf{89}(2004)023005.

\bibitem{18} Herrera, L.: Int. J. Mod. Phys. D \textbf{20}(2011)1689.

\bibitem{z1} Sharif, M. and Yousaf, Z.: Eur. Phys. J. C \textbf{75}(2015)58.

\bibitem{15} Kanno, S., Sasaki, M. and Tanaka, T.: Prog. Theor. Exp.
Phys. \textbf{2013}(2013)111E01.

\bibitem{16} Geng, W. and L\"{u}, H.: Phys. Rev. D \textbf{90}(2014)083511.

\bibitem{17} Sharif, M. and Bhatti, M.Z.: Mod. Phys. Lett. A
\textbf{29}(2014)145019; ibid. \textbf{29}(2014)1450094; ibid.
\textbf{29}(2014)1450165; Phys. Scr. \textbf{89}(2014)084004.

\bibitem{19} Zannias, T.: Phys. Rev. D \textbf{41}(1990)3252.

\bibitem{20} G\'{o}mez-Lobo, A.G.P.: Class. Quantum Grav.
\textbf{25}(2008)015006; Herrera, L., Ospino, J., Di Prisco, A.,
Fuenmayor, E. and Troconis, O.: Phys. Rev. D
\textbf{79}(2009)064025.

\bibitem{21} Springel, V. et al.: Nature \textbf{435}(2005)629.

\bibitem{21a} Lambiase, G. Phys. Rev. D \textbf{90}(2014)064050.

\bibitem{22} Ellis, G.F.R.: Gen. Relativ. Gravit. \textbf{41}(2009)581.

\end{thebibliography}
\end{document}